# VISUAL CHECKING OF SPREADSHEETS

Ying Chen and Hock Chuan Chan
School of Computing, National University of Singapore, Singapore
(Contact author: Hock Chan, email: chanhc@comp.nus.edu.sg)

**ABSTRACT**

*The difference between surface and deep structures of a spreadsheet is a major cause of difficulty in checking spreadsheets. After a brief survey of current methods of checking (or debugging) spreadsheets, new visual methods of showing the deep structures are presented. Illustrations are given on how these visual methods can be employed in various interactive local and global debugging strategies.*

## 1. INTRODUCTION

Spreadsheets are easy to use and very hard to check. Large percentages of spreadsheets in actual use have been found to contain errors, and users do not have effective means of checking spreadsheets. The main difficulty that users have in understanding spreadsheets is the difficulty of establishing cell connections - how each cell depends on other cells. These connections form a structure. The usual tabular layouts for texts and numbers do not show the actual connections. They may even suggest a wrong structure. The difference between the surface structure (what the cell arrangements and values suggest to the user) and the deep structure (what the cell connections really are, as described in the formulas) can add to the difficulty in understanding spreadsheets.

Some systematic ways of creating spreadsheets have been recommended [1]. This approach specifies some design requirements so that the spreadsheet will be easier to maintain and be understood by someone other than the developer. The possibility of making errors may be reduced. However, errors can still occur, even if users are generally careful. Spreadsheet checking is also necessary at the development stage.

Currently, deep structures can be understood only by examining formulas one by one. This process is extremely tedious and error prone. Researchers are trying to make the deep structures visible, through visual drawings. Various visual tools and strategies for using these tools are presented, so as to help spreadsheet users understand their spreadsheets faster and more thoroughly.

## 2. SPREADSHEET STRUCTURES

Cells in a spreadsheet are organised as tabular form. Related cells are often put together. Naturally, users deduce the spreadsheet structure through relative positions of cells as they visually cluster them together. The spreadsheet structure is obtained through the visual scanning of cells. So it is called visual/surface structure. However, spreadsheet calculations are based on the formulae in cells. Cells are connected through formulae and form another kind of structure, called computational/deep structure. It reflects data flows in a spreadsheet.

Table 1(a) shows a spreadsheet with some values. The "visual" relationships among cells, identified by users through the relative spatial positions of cells in the spreadsheet provide a surface structure. At first sight, a user may form an impression that column A is collapsed into column B, column B into column C and column C into column D. The dotted arrows represent the supposed surface level of this spreadsheet. Table 1(b) shows the deep structure, with formulas determining the actual cell connections. An arrow goes from a precedent cell to a dependent cell (a dependent cell's formula refers to a precedent cell). This structure is different from table 1(a). The inconsistency between surface level and deep level can happen in two major ways 2: (1) References in a single formula is non-unitary and non-organised, as in E6 * F9 * A5 * C4; and (2) References in several formulas are interwoven in "the cell dependency network".

Table 1(a). The surface structure

Table 1(b). The deep structure

**Table1. the structures of a spreadsheet**

It is difficult to understand a spreadsheet when the two structures in the spreadsheet are very inconsistent, because information search is controlled by the surface structure. The memory load that is used to extract the deep structure will increase largely [3]. As a consequence, users will spend more time on learning and memorizing the deep structure than doing spreadsheet calculations.

This suggests that surface structure could be used effectively as mnemonic aids if the layout allows the direct spatial mapping between the elements of surface structure and deep structure. Some rules can be used to create well-organised formulas (the links in networked formulas are not crossed and their direction are uniform). Then the memory load goes down.

It is worth noting that some specialized spreadsheet programs have been developed to make the deep structures explicit during the model development stage. For example, Spreadsheet 2000 (http://www.emer.com/s2k/) avoids the typical tabular spread of cells, and thus totally avoids giving any surface structure. Models are built with predefined metaphors, such as operators, grids, charts and forms. This avoids the problem of surface structure misunderstanding. Another software, Storeys (http://www.profunda.com), allows users to built 3-dimensional spreadsheets in a more explicit manner, compared to the usual 2-dimensional worksheets. This again avoids to some degree the surface structure of a 2-dimensional spreadsheet when the model is really 3-dimensional.

Given that the most popular spreadsheets (from VisiCalc to Lotus 1-2-3 to Microsoft Excel) adopt the general tabular format without explicit structures, a very important approach is to dispel the surface structure by displaying the deep structure directly. The visibility of deep structures can help spreadsheet users put most of their attention to checking rather than memorizing the structures.

Structures can be viewed at two levels of granularity: Cell and Module. At cell level, the basic elements in the structure are cells. The structure represents the relationships among cells. Any change in the structure is reflected at cell level. In contrast, the module level is used to describe spreadsheets at the macro level. A module is made up of a set of cells that have the same function. The structure describes relationships among modules. It is a high level view of how a spreadsheet model operates.

At module level, each module element is a block of cells. The definition of a block at surface level is different from that at deep level. At deep level, a block is defined by its function, e.g., Input, Output, Decision, Parameter and Formulae [1], or simply Input, Process and Output. The relationships among the blocks represent the logic flows in a spreadsheet model. In contrast, at surface level, blank cells divide a spreadsheet into different blocks. The cells in a block may not represent a separated function of a spreadsheet. However, the spatial position of a block may represent corresponding functions: Input part is usually put on the upper or at the left of a spreadsheet while Process part is put in the middle and Output part is put at the bottom or the right.

The same problem of inconsistency between surface and deep level exists at module level as well as at cell level. The usual method to solve this problem at module level is to follow the requirements of structured design: Put all relevant cells into the corresponding spatial position of one functional part.

Accordingly, debugging tools should let users check spreadsheets locally (at cell level) and globally (at module level) so that errors can be detected more easily.

## 3. CURRENT METHODS

This section reviews the current dominant methods to find errors in spreadsheets.

**Presentation**

Lacking good visualization tools, presentation is an obvious method for manual error detection. Spreadsheets can be shown on-screen, or printed on paper, and with or without formulas. Experiments suggested that spreadsheets are better understood with formulae on paper than with numbers on screen [4]. However, detection rates for all these presentation methods are very low.

**Program Code inspection**

Theories from error detection in program codes were adopted to discover errors in spreadsheets [5]. However, spreadsheets are quite different from program codes. Firstly, program codes are represented in a one-dimension way while cells in spreadsheets are listed in a two-dimensional form. Secondly, the models (formula and deep structure) of a spreadsheet are buried in the surface data, unlike a program where program codes are organised according to some kinds of semantic and syntax. Some restrictions are imposed on the program codes to make them meaningful and easy to be understood. Thirdly, few organizations have comprehensive policies for spreadsheet development while software engineering has built good rules for programmers' use. The development and detection phases are so tightly correlated that good development will make detection far easier. So the ways to detect errors in programs cannot be completely applied to error detection in spreadsheets. The average error detection rate for individuals is 63% [5]. Some errors are still left in spreadsheets. It is necessary to develop more suitable ways to detect errors in spreadsheet.

**Spreadsheet Analysis Reports**

A number of software packages have been developed to analyse spreadsheets and present reports, mainly for auditing purposes. Early products, such as Cambridge Spreadsheet Analyst and Spreadsheet Auditor, to help analyse spreadsheets have appeared since 1986[6]. More recent products include the following. The Spreadsheet Detective (http://www.gg.net.au/detective/) analyses a spreadsheet and adds cell shadings and comments to help users identify errors. Operis Analysis Kit (http://www.operis.com/) is another add-in for Excel 97, and helps to identify various kinds of spreadsheet errors. Spreadsheet Professional (http://www.eastern-software.com/) also provides analysis to help identify spreadsheet errors.

**Interactive tools**

Recent technological advances allow researchers to visually explore the discrepancy between surface and deep structures of spreadsheets. Tools have been developed to help users identify data flows at cell and module levels. Davis [7] described two interactive tools for auditing spreadsheets: an on-line flowchart-like diagram[1] and a tool which represents "cell dependency network" [3] by drawing arrows cells on the spreadsheet. Both tools were found to be helpful in investigating cell dependencies and debugging. Both tools show better error detection than the traditional screen / paper presentations. The arrow tool was the best. It was also found that users prefer visual interactive tools rather than paper printouts or screen scanning [8].

At a higher level, Isakowitz et al[9]. Developed an algorithm to extract logic structures from spatial layouts automatically. Four principal components are identified from the original spreadsheet: schema, data, editorial and binding. Then users can easily construct logic flows from schema. It is an easier way to check spreadsheets. Users can debug spreadsheets with schema and data and avoid cell-by-cell checking. One drawback related to this technology is that there is no direct mapping between spreadsheets and schema. Users need to try to recognise the corresponding components of schemas in spreadsheets. So these schemas impose a heavy memory load on users.

To make interactive tools more convenient, Igarashi et al.[10] described ways to visualise spreadsheet structures at varying levels of connections. Initially, first-level precedents and dependents of a cell are shown. Subsequently, the entire structure of a spreadsheet related to one cell is displayed. Furthermore, displays can be static or animated. However, it has some limits: spreadsheets cannot be checked area-by-area or module by module; all precedents or dependents of a cell cannot be obtained at once and display is based on only one cell at a time. Microsoft Excel spreadsheet provides the same functionality in its set of auditing tools. Various levels of precedents and dependents for a selected cell can be displayed, with arrows.

## 4. NEW VISUAL METHODS

A set of new visual tools for understanding spreadsheets is briefly presented here. These tools help to highlight different functional parts, data flows and levels of a spreadsheet. These may allow users to debug their spreadsheets with proper strategies. These have been implemented in Excel Visual Basic for Applications, and can be added in to any Excel installation.

*1. Functional Identification Tool*

All cells in a selected region are classified into four kinds: Input, Output, Processing and Standalone. Input cells have dependents but no precedents while output cells have precedents but no dependents. Cells with precedents and dependents belong to processing parts. Standalone cells have neither precedents nor dependents. Different colours are used to fill these four types of cell so that users can see the function of cells immediately.

*2. Multi-precedents and Dependents Tool*

The multi-precedents/dependents tool shows the precedents/dependents of all cells in a selected block. It can illustrate relationships among cells beyond one level at one click.

*3. Block-precedents Tool*

This defines a set of similar cells as a block. Their precedents in different neighbouring areas make up different blocks. This block-precedents tool shows the relationships wrong blocks, not cells. One level is added at one click.

*4. In-block-precedents-dependents Tool*

In-Block-Precedents-Dependents shows cell connections within a selected region, without incoming and outgoing arrows connected to this region. It can be applied when users are only concerned with cell relationships in a particular region. All distractions outside this region can be avoided.

*5. Separated-blocks Tool*

This tool identifies groups of connected cells. There are no connections across groups. Different cell and arrow colours are used to mark different groups. Each group represents a separate model in the worksheet. So different models are easily identified.

*6. Level Label Tool*

Level label assigns a level number to cells in a selected block in terms of the longest distances between the cells and input cells. Input cells can be checked first while output cells can be checked last. The order to check contents in cells can help to minimise repetition and disorderliness.

## 5. STRATEGIES

A new tool should ideally be coupled with a set of strategies that defines the best way to use it. These strategies could be developed by experts familiar with the new tools, and can form a basis for other users to apply the new tools in their understanding of spreadsheets and detection of errors. Some strategies for using the new tools are presented below.

Understanding a spreadsheet relies on a good understanding of the "Cell dependency network". Data effects flow from the input to the output cells. Due to the flow direction, "Cell dependency network" can be regarded as a directed graph. The value of a cell in the flow can only be affected by its precedents. So if users are sure where the problem falls or which functional parts the problem belongs to, users can concentrate on a specific area and local debugging strategies are enough. However, if users have no idea about what their spreadsheets do, global debugging strategies are more suitable to help users establish thorough debugging steps. Next, these two debugging strategies are illustrated.

|   | A | B |
|---|---|---|
| 1 | Revenue | 30000 |
| 2 | Cost | 20000 |
| 3 | Profit b/f Tax | 10000(B1-B2) |
| 4 | Tax | 5000 |
| 5 | Profit after Tax | 5000(B3-B4) |

(a)

|   | A | B |
|---|---|---|
| 1 | Revenue | 30000 |
| 2 | Cost | 20000 |
| 3 | Profit b/f Tax | 10000 |
| 4 | Tax | 5000 |
| 5 | Profit after Tax | 5000 |

(b)

|   | A | B |
|---|---|---|
| 1 | Revenue | 30000 |
| 2 | Cost | 20000 |
| 3 | Profit b/f Tax | 10000 |
| 4 | Tax | 5000 |
| 5 | Profit after Tax | 5000 |

(c)

|   | A | B |
|---|---|---|
| 1 | Revenue | 30000 |
| 2 | Cost | 20000 |
| 3 | Profit b/f Tax | 10000 |
| 4 | Tax | 5000 |
| 5 | Profit after Tax | 5000 |

(d)

|   | A | B |
|---|---|---|
| 1 | Revenue | 30000 |
| 2 | Cost | 20000 |
| 3 | Profit b/f Tax | 10000 |
| 4 | Tax | 5000 |
| 5 | Profit after Tax | 5000 |

(e)

**Table 2. An example of a broken link**

**Local debugging strategy I: Broken Link**

One common error occurs when a user accidentally types a value instead of a formula. It is very difficult to catch because the value in this cell is correct, until such time when the values of precedents of this cell are changed. Two methods can be used to catch this type of error: functional identification and in-block-precedents-dependents.

For example, table 2(a) shows a simple spreadsheet. Table 2(b) and 2(d) show the correct visual result after applying the functional identification tool or the in-block precedent dependent tool. If a user accidentally enters a value instead of a formula in cell 133, then applying the same tools will result in the visual effects shown in table 2(c) and 2(e). The pink coloured cells (cells Bl and B2) in table 2(c) alert the users to errors, since pink coloured cells are standalones, and not used by other cells. In a well-structured spreadsheet, all cells should be connected. Some exceptions could occur, such as in column headings. So it is easy to find that the correct form of 133 should be B1 - B2 and not 10000. Table 2(e) shows a shortened chain of arrows, starting from the middle of the model. Hence, it is easy for users to spot this error.

**Local debugging strategy II: Unwanted Link**

Due to similarity within a set of formulas, relationships among several sets of cells may be regular. Any irregularity indicates possible error. Multi-precedents/dependents may help users to catch irregularities. For example, the wrong cell reference of C4 in Table 3(a) is easily caught by applying multi-precedents to the block of cells C2 to CS. A diagonal downward arrow is easily spotted.

In Table 3(b), users find that values for % change for different regions contribute to the overall % change after multi-dependents is applied to cells B4 to B7. From logical consideration, it is somewhat strange since interest rates are seldom aggregated. So the logical error is caught.

|   | A | B | C |
|---|---|---|---|
| 1 | Deposit | Rate | Payment |
| 2 | 3000 | 0.03 | 90 |
| 3 | 5000 | 0.035 | 175 |
| 4 | 8000 | 0.05 | 280 |
| 5 | 10000 | 0.06 | 600 |

(a)

|   | A | B | C | D |
|---|---|---|---|---|
| 1 | Regional Sales & Profits |  |  |  |
| 2 |  | 1981 | 1980 | % Change |
| 3 | Sale($'000) |  |  |  |
| 4 | East Asia | 79000 | 60755 | 16.2% |
| 5 | Middle East | 12321 | 15234 | 19.1% |
| 6 | Europe | 3565 | 3200 | 11.4% |
| 7 | Caribbean | 43123 | 125954 | 11.2% |
| 8 |  | 138897 | 125954 | 19.7% |

(b)

Table 3. Examples of Unwanted Links

**Local debugging strategy III: Localised Links**

A spreadsheet may be large and complicated. To check many cells in a large region at a time may include many irrelevant details. One strategy is to examine the links within a small selected region. Users can use the in-block-precedents-dependents tool to delve into details of a selected block, not distracted by connections to cells outside the block.

**Global debugging strategy I: Localization**

Local checking can allow users to delve into the details. However, users may want to know the data flow at a higher level or do not want to miss any part of a spreadsheet. Global checking strategies may help users identify the logical structure or to plan thorough debugging steps. The first strategy helps user identify sub-regions within a region.

In the example in table 4, which shows a correct model, the functional identification tool is first applied to identify input, output, processing and standalone parts. The light green, dark green and grey cells show the different cell functions. Users can have an initial impression of the spreadsheet structures. The functional modules of the spreadsheets are also identified by the different colour regions. Output modules are to the right and bottom of the spreadsheet (most of column H, and rows 20 and 21 in the table); input modules occupy the centre (columns B to G, in rows 5 to 9, and 13 to 16); and processing modules are in two rows (row 10 and row 17 in the table).

|    | A | B | C | D | E | F | G | H |
|----|---|---|---|---|---|---|---|---|
| 2  | **Fixed Assets** | | | | | | | |
| 3  | Cost | Freehold Buildings | Leasehold property | vehicles | Equipment | Computers | Electrical fittings | Total |
| 4  |   | $ | $ | $ | $ | $ | $ | $ |
| 5  | Balance at 1.4.80 | 2,105,165 | 52,422,655 | 2,500,444 | 880,355 | 2,000,500 | 120,500 | 60,029,619 |
| 6  | Additions | 188,367 | 1,826,545 | 388,567 | 455,676 | 380,600 | 1,200,500 | 4,440,255 |
| 7  | Disposals | | | (55,650) | (75,800) | (205,900) | (100,800) | (438,150) |
| 8  | Transfer to other assets | | | | | (9,050) | | (9,050) |
| 9  | Translation difference | 23,156 | | 12,565 | 12,885 | 12,565 | | 61,171 |
| 10 | Balance at 31.3.81 | 2,316,688 | 54,249,200 | 2,845,926 | 1,273,116 | 2,178,715 | 1,220,200 | 64,083,845 |
| 11 | | | | | | | | |
| 12 | **Accumulated Depreciation** | | | | | | | |
| 13 | Balance At 1.4.80 | 300,150 | 800,500 | 1,500,890 | 410,985 | 1,234,567 | 110,921 | 4,358,013 |
| 14 | Depreciation for the year | 105,000 | 650,500 | 350,900 | 123,456 | 123,321 | 112,344 | 1,465,521 |
| 15 | Released on disposals | | | (45,000) | (40,500) | (199,288) | (100,855) | (385,643) |
| 16 | Translation difference | 8,650 | | 9,520 | 3,755 | 9,012 | | 30,937 |
| 17 | Balance at 31.3.81 | 413,800 | 1,451,000 | 1,016,310 | 497,696 | 1,167,612 | 122,410 | 5,468,828 |
| 18 | | | | | | | | |
| 19 | Net book value | | | | | | | |
| 20 | At 31.3.81 | 1,902,888 | 52,798,200 | 1,029,616 | 775,420 | 1,011,103 | 1,097,790 | 58,615,017 |
| 21 | At 1.4.80 | 1,805,015 | 51,622,155 | 999,554 | 469,370 | 765,933 | 9,579 | 55,671,606 |

**Table 4. Localization**

Users can use the multi-precedents tool to check similar output cells to detect any irregularity or critical output cells to identify data flows. The arrows in table 4 are the results of applying the multi-precedents tool to B20 to G20. The logical flows in B5 to G20 can be easily identified at a glance. More importantly, this shows that each column is in effect independent. Column values are summed in column H, but otherwise they are independent. Thus, the overall region has been decomposed into single columns. An alternative tool to use is the block-precedents tool. There are fewer arrows, providing a clearer view.

**Global debugging strategy II: Separation**

Table 5 illustrates the strategy of separation. Unrelated groups of cells are identified using the separated-blocks tool. Different colour arrows visually identify four models in the worksheet. There are no connections among them and no standalone parts. All cells are properly connected.

With the identification of separated groups, the user can now focus on each group without being distracted by other cells. In some cases, separated groups mean errors if the groups are not meant to be separated. For example, some links could be missing.

**Global debugging strategy III: Stratification**

The arrows linking an output cell and all of its precedents form a tree. The output cell is correct if the formula and the first level precedents of this cell are correct. It is same for all cells in a spreadsheet. So if a cell is checked after all of its precedents have been checked, the repetition of checking can be avoided. The checking efficiency is improved. Hence, all cells are stratified with the level label tool. All cells are assigned to a level number according to the longest distance from them to the input cells. In Table 5, the corresponding level label for each cell is given. Checking should proceed from level one onward. In this sequence, output cells are checked last while input cells are check first. Then users have a visual plan to examine the spreadsheet level by level without repetition.

## 6. CONCLUSION

Spreadsheet errors are common, and can have serious decision consequences. In order to help users understand spreadsheets and identify errors in a visual manner with lower cognitive loads, various local and global strategies with a new set of interactive tools are presented. Local debugging strategies help users delve into details while global debugging strategies help users establish thorough and more systematic debugging steps.

Compared to the earlier error detection methods, this new set of tools can make deep structures more visible and obvious by the arrows connecting precedent and dependent cells. The tools provide more convenience by enabling all precedents and dependents of a group of cells to be shown at once, instead of requiring many clicks from the user. Cell relationships can also be limited to a selected region, so as to allow more focused understanding and detection of errors. Visual comprehension is facilitated through the use of different colours to represent different functional parts in a spreadsheet. Visual aids by the interactive tools can make users concentrate on the important task of verifying the model and not on remembering and recalling cell connections.

|    | A                               | B        | C        | D        | E       |
|----|---------------------------------|----------|----------|----------|---------|
| 1  | **Company Balance Sheet December 31, 1981** |          |          |          |         |
| 2  |                                 |          | 1981     |          | 1980    |
| 3  |                                 |          |          |          |         |
| 4  | Share Capital and Reserves      |          |          |          |         |
| 5  | Share capital                   | ① 45000  |          | ① 45000  |         |
| 6  | Capital reserves                | 99200    |          | 10000    |         |
| 7  | Retained profits                | 17768    |          | 35788    |         |
| 8  |                                 |          | ② 161968 |          | ② 90788 |
| 9  | Represented by:                 |          |          |          |         |
| 10 | Fixed Assets                    |          |          |          |         |
| 11 | Other Investment                |          | ① 268241 |          | ① 9846  |
| 12 | Deferred Taxation Asset         |          | 22026    |          | 7667    |
| 13 | Current Assets                  |          |          |          | 22303   |
| 14 | Stocks                          |          |          |          |         |
| 15 | Trade debtors                   | ① 13593  |          | ① 9996   |         |
| 16 | Other debtors and deposits      | 3822     |          | 423      |         |
| 17 | Fixed deposits                  | 6353     |          | 102882   |         |
| 18 | Cash                            | 152748   |          | 100404   |         |
| 19 | Total current assets            | 14996    |          | 5386     |         |
| 20 | Less current liabilities        | ② 191512 |          | ② 219091 |         |
| 21 | Trade creditors                 |          |          |          |         |
| 22 | Provision for claims            | ① 108593 |          | ① 10221  |         |
| 23 | Provision for upgrades          | 10960    |          | 74963    |         |
| 24 | Fuel price equalization         | 5604     |          | 14691    |         |
| 25 | Loan                            | 53127    |          | 52227    |         |
| 26 | Income tax payable              | 72000    |          | 4000     |         |
| 27 | Proposed dividends              | 12000    |          | 129      |         |
| 28 | Total current liabilities       | 13048    |          | 2700     |         |
| 29 | Net Current Assets              |          | ③ 275332 | ② 158931 |         |
| 30 | Less: Non-current liabilities   |          | ⑤ -83820 |          | ③ 60160 |
| 31 | Provision for replacement       |          |          |          |         |
| 32 | Provision for service benefits  | ① 3442   |          | ① 4188   |         |
| 33 | Deferred taxation liabilities   | 28857    |          | 5000     |         |
| 34 | Total non-current liabilities   | 12180    |          | 0        |         |
| 35 |                                 |          | ② 44479  |          | ② 9188  |
| 36 | Net Assets                      |          | ③ 161968 |          | ③ 90788 |

**Table 5. Separation and Stratification**